\documentclass[prl,twocolumn,showpacs,preprintnumbers,amsmath,amssymb]{revtex4}

\usepackage{graphicx}
\usepackage{dcolumn}
\usepackage{bm}

\begin{document}


\title{ Theory of  superconducting fluctuations
 in the thinest  carbon nanotubes }

\author{Krzysztof Byczuk }
\affiliation{\centerline{Theoretical Physics III, Center for Electronic Correlations and 
Magnetism, Institute for Physics,} 
\centerline{University of Augsburg, D-86135 Augsburg, Germany, }}
\affiliation{\centerline { Institute of Theoretical Physics,
Warsaw University, ul. Ho\.za 69, PL-00-681 Warszawa, Poland, }}

\date{\today}
\begin{abstract}

The low-energy electronic  Hamiltonian for the thinest 
zigzag carbon nanotube, embedded into a dielectric host, is derived 
and its phase diagram is discussed.
The specific multi-band structure and the microscopic form 
of the electron-electron interaction in this systems is considered.
The interband repulsive interaction, which is almost unscreened, 
leads to a retarded intraband attraction between the electrons and can stabilize the Cooper pairs.
For a dielectric constant of the host $\epsilon_d\sim 2-4$, the theory predicts that 
the superconducting fluctuations should develop, which is  in agreement with the experiment.
For $\epsilon_d \gtrsim 8$, the density wave fluctuations should be amplified. 
Between the two phases, there is a metallic state where all two-particle 
fluctuations are suppressed.

\end{abstract}

\maketitle


Carbon nanotubes \cite{Iijima91} 
attract the attention  because of their unique 
geometrical structure, mechanical, chemical  or electrical properties,   and possible applications in 
electronic devices \cite{Review,Saito98}.
A single-wall carbon nanotube (SWNT) is made of a graphite layer rolled up into a cylinder \cite{Saito98} 
with the smallest diameter  about $4\AA$ \cite{Qin00}. 
In such thinest SWNTs, superconducting fluctuations (SCFs) were observed 
with the  mean-field critical temperature  about $15\;K$ \cite{Tang01}.
Since it is a one-dimensional (1-d) system, the true off-diagonal long-range order is suppressed.
Nevertheless, the SCFs might develop and the  question arises about  their microscopic  origin. 

The low-energy electronic excitations in the metallic SWNT with a moderate diameter are described by the 
Luttinger liquid theory with four collective modes \cite{Egger97}.
The SCFs induced by the backscattering of the electrons might appear only at the lowest
energy scale and  it is rather unlikely that they can be ascribed to 
 the occurrence of such fluctuations in the 
$4\AA$-SWNTs \cite{Tang01}.
Theories considering  lattice vibrations  require a large 
reduction of the the repulsive electron-electron interaction to stabilize the Cooper pairs \cite{Gonzalez02}.
The renormalization group analysis confirms 
that the SCFs due to phonons in the SWNTs can  easily be destroyed by the competing
 Coulomb repulsion \cite{Sedeki02}.
The theory \cite{Gonzalez02}  explains the superconductivity in 
ropes of the SWNTs \cite{Kociak01}, where also the intertube hopping plays the role.
However, in the experiment  \cite{Tang01},
the SWNTs were embedded into a zeolite dielectric host and they were isolated from each others.
Therefore, it seems rather unlikely that the SCFs in such a system may  develop 
due to  phonons.

In the present paper we derive a low-energy electronic Hamiltonian  for 
the thinest zigzag SWNTs embedded into a dielectric host. 
Because of a large curvature of the cylindrical shell in such SWNTs, there are three 1-d bands 
crossing the Fermi level.
We find  a phase diagram determined by the slowest decaying in space two-particle correlator. 
For realistic parameters, corresponding to the experimental realization \cite{Tang01},
the SCFs are developed in the system.
They are induced by the unscreened interband Coulomb interaction.
Changing the dielectric properties of the host, other possible phases are predicted.

The simplest tight-binding approximation predicts that the $(n,m)$ SWNT is metallic if ${\rm mod}(n-m,3)=0$ (where
$(n,m)$ is a chiral vector describing uniquely the infinitely long SWNT) \cite{Saito98}.
In such a case there are two  bands crossing the Fermi level. 


When the curvature  of the cylinder
is not neglected, it turns out that from all of the metallic  SWNTs only 
those with the $(n,n)$  chiral vectors (armchair SWNTs)
remain metallic and the rest are semiconductors with a small gap in the spectrum \cite{gap}. 
However, for the thinest SWNT $(n,0)$ (zigzag SWNTs) 
with $n=4,5$ and $6$, it has been shown that 
the Koster-Slater parameterization in the tight binding approximation  is not adequate
because it does not capture correctly the hybridization between the $\sigma-\pi$ orbitals \cite{Blase94}. 
The functional density calculations within the local-density approximation (LDA)
 have shown that such nanotubes are metallic. 
Because of the very strong $\sigma-\pi$ hybridization, the nondegenerate 
band originating from a $\Gamma-M$ line in the 
hexagonal Brillouin zone is strongly repulsed and pushed down
crossing the two-fold degenerate valance bands. 
As a result the Fermi level is  lowered and there are three 1-d bands that are partially populated by 
the electrons leading to the metallic band structure. 

The low-energy theory for the thinest zigzag SWNTs is given by {\em the three-band Tomonaga-Luttinger 
 model with one nondegenerate band and two-bands that are two-fold degenerate}. 
The corresponding 1-d Hamiltonian has the following kinetic part
\begin{equation}
H_0=-i \sum_{j=0}^2\sum_{\sigma , r} (-1)^r v_F^j\int dx \;
\psi^{\dagger}_{rj\sigma}(x)\partial _x\psi_{rj\sigma}(x),
\label{1}
\end{equation}
where $\psi_{rj\sigma}(x)$ are  slowly varying Fermi fields
 for the right $r=R$ and the left $r=L$ moving electrons
with the spin $\sigma$ in one of the three bands labeled by $j$. 
The non-degenerate band is labeled by $j=0$ and has  the Fermi velocity $v_F^0$.
The two-fold
degenerate bands are labeled by $j=1$ and $2$ and have the same Fermi velocity $v_F$. 
The Fermi velocities might be determined  from the LDA band structures,
which were calculated in 
the supercells geometry of the SWNTs \cite{Blase94}.
In the experimental realization \cite{Tang01}, the SWNTs were in  1-d channels of a porous zeolite 
(AlPO$_4$-$5$)  and, in principle, 
interacted with the walls of the host by  the van der Waals forces. 
The qualitative structure of the bands should not be changed with respect to the LDA results, however,
 small differences in the position 
of the Fermi level and the bottom of the $\Gamma-M$ band might be expected.
Therefore, the numerical values for the Fermi velocities require more detailed
investigation considering  the presence of the host.
In the following we keep these two velocities as parameters in the model.
For further convenience  we introduce the ratio $\eta\equiv v_F^0/v_F$ and parameterize the 
model by $v_F$ and $\eta$.

The long-range Coulomb interaction between the electrons is modified  by the 
zeolite host, which has a dielectric constant $\epsilon _d $.
 One-dimensional wire embedded in a host with a different dielectric constant was studied in \cite{Byczuk99}
where the potential describing the electron-electron and the electron-image-charge interactions was
derived.
Using now the wave functions centered at the cylindrical shell of the SWNT, we calculate the 
Fourier component $V(q)$ of the electrostatic potential in the tube embedded into the 
dielectric host.
Explicitly, the potential takes the form
\begin{eqnarray}
V(q)=\frac{e^2}{\pi \epsilon_1}
\left[I_0(qR) K_0(qR)+\right. \nonumber \\ 
\left. \frac{\left(1-\frac{\epsilon_d}{\epsilon_1}\right)K_1(qa)K_0(qa)}{\frac{\epsilon_d}{\epsilon_1}
K_1(qa)I_0(qa)+K_0(qa)I_1(qa)}
I_0(qR)\right],
\label{potential}
\end{eqnarray}
where $a$ is the radius of the cavity in the zeolite, $R\leq a$ is the radius of the SWNT, 
$\epsilon_1$ is the dielectric constant of the SWNT,  and $I_n$ and $K_n$ are modified Bessel functions.
For small $q$ the potential diverges as $\ln|qR|$.
For large  $q$ the tail of the  potential $V(q)$ is
suppressed because of the image charges.
Therefore, a scattering with large momentum transfer, in particular the backscattering with $q=2k_F$, in the 
SWNTs embedded into the dielectric host  should be
 suppressed  and change the phase diagram only at the lowest energy/temperature scales.

The low-energy Hamiltonian requires the screened electron-electron Coulomb potential
obtained by removing self-consistently high-energy degrees of freedom. 
In the random-phase-approximation  the particle-hole polarization diagrams are only kept and 
it leads to the dynamically screened electron-electron Coulomb potential given by the algebraic equation
$
\sum_{ll'}\epsilon_{ll'nn'}(q,\omega)V^s_{ll'mm'}(q,\omega)=V_{nn'mm'}(q),
$
where $V_{nn'mm'}$ ($V^s_{ll'mm'}$) is the matrix element for the bare (screened) Coulomb potential and 
$\epsilon_{ll'nn'}=\delta_{ln}\delta_{l'n'}-\Pi_{ll'}V_{ll'nn'}$ is the matrix dielectric function
 expressed by the
particle-hole polarization function $ \Pi_{ll'}(q,\omega)$ \cite{Tavares01}. 
The indexes correspond to the Bloch functions of the corresponding band. 
In the static limit $\omega=0$ and for the momentum transfer $q\rightarrow 0$ we obtain 
the screened intraband  potential  
$V^s_{0000}\approx V_{0000}(1-2 V_{1111}\Pi_{11})/\epsilon_{intra}$ 
and $V^s_{1111}=V^s_{2222}\approx V_{1111}(1-V_{0000}\Pi_{00})/\epsilon _{intra}$
with $\epsilon _{intra}\approx (1-V_{0000}\Pi_{00})(1-2V_{1111}\Pi_{11})$ \cite{comment0}.
However, the interband Coulomb potential remains almost unscreened within this limit, i.e 
$V^s_{0101}=V^s_{0202}\approx V_{0101}/\epsilon _{inter}$ with $\epsilon _{inter}\approx 1$.

The effective interaction  between the electrons is described by the two-particle  part of the Hamiltonian
\begin{equation}
H_F=\frac{1}{4}
\sum_{i,j=0}^2\sum_{rr'}\sum_{\sigma\sigma'}\int dx dx' \rho_{ri\sigma}(x)U_{ij}(x-x')\rho_{r'j\sigma'}(x'),
\label{2}
\end{equation}
where $\rho_{ri\sigma}(x)=\psi^{\dagger}_{rj\sigma}(x)\psi_{rj\sigma}(x)$ is the density fluctuation
operator. 
Taking the bare  potential 
$V_{iiii}(q)=V_{ijij}(q)=V(q)$
and approximate it by its average value $V_{\rm av}$ calculated between  the infrared-cutoff 
$q_c^{\rm inf}= 2\pi/L$, 
where $L$ is the nanotube length, and the ultraviolet-cutoff $q_c^{\rm ult}=5\cdot 10^8\;m^{-1}$ (which 
gives the energy cutoff $E_c=0.1\; eV$),
 we obtain  the following 
local potential $U_{ij}(x-x')=U_{ij}\delta(x-x')$ with
$U_{00}=V_{\rm av}/[1+2V_{\rm av}/\pi\eta v_F]\equiv U_0 $,
$U_{11}=U_{22}=U_{12}=V_{\rm av}/[1+4V_{\rm av}/\pi v_F]\equiv U_1 $, and
$U_{01}=U_{02}=V_{\rm av}\equiv U' $ \cite{comment1}.

To find exponents in the one- and two-particle correlation functions in the 
thinest SWNT, we need the explicit solution of the 
Hamiltonian $H=H_0+H_F$, where $H_0$ and $H_F$ are expressed  by Eqs. (\ref{1}) and (\ref{2}). 
This can be achieved by using the nonabelian 
bosonization where  conjugate fields $\phi_{j\sigma}(x)$ and $\Pi_{j\sigma}(x)$ are introduced
for each of the band $j$ and the spin $\sigma$ \cite{bosonization}. 
The Hamiltonian separates into different sectors if we introduce the collective variables
\begin{equation}
\left(
\begin{array}{c}
\phi_{0c}\\
\phi_{0s}
\end{array}
\right)=\frac{1}{\sqrt{2}}\left(
\begin{array}{rr}
1&1\\
1&-1
\end{array}
\right)
\left(
\begin{array}{c}
\phi_{0\uparrow}\\
\phi_{0\downarrow}
\end{array}
\right)
\label{3}
\end{equation}
and 
\begin{equation}
\left(
\begin{array}{c}
\phi_{c}\\
\phi_{s}\\
\phi_{f}\\
\phi_{x}
\end{array}
\right)=\frac{1}{2}\left(
\begin{array}{rrrr}
1&1&1&1\\
1&-1&1&-1\\
1&1&-1&-1\\
1&-1&-1&1
\end{array}
\right)
\left(
\begin{array}{c}
\phi_{1\uparrow}\\
\phi_{1\downarrow}\\
\phi_{2\uparrow}\\
\phi_{2\downarrow}
\end{array}
\right),
\label{4}
\end{equation}
for nondegenerate and degenerate bands, respectively.
There are six types of the collective oscillations in the system: four of them are neutral 
and two of them are charged.
This feature distinguishes the thinest zigzag SWNTs from the other metallic SWNTs
with moderate diameters.

The interaction appears only between the charged density modes. 
Rescaling the fields $\tilde{\phi}=\phi/\sqrt{K}$ and $\tilde{\Pi}=\sqrt{K}\Pi$,  we  obtain the 
following Hamiltonian in the charge sector
\begin{eqnarray}
H_c=\int dx \left(
\frac{\tilde{v}_{Fc}^0}{2}
\left\{ \tilde{\Pi}_{0c}^2(x) +\left[ \partial_x \tilde{\phi}_{0c}(x)\right]^2\right\}+ \right. \nonumber \\
\left. \frac{\tilde{v}_{Fc}}{2} \left\{ \tilde{\Pi}_{c}^2(x) + 
\left[ \partial_x \tilde{\phi}_{c}(x)\right]^2\right\}+
\lambda  \left[\partial_x \tilde{\phi}_{0c}(x) \right]\left[\partial_x \tilde{\phi}_{c}(x) \right] \right),
\label{5}
\end{eqnarray}
where the Luttinger liquid parameters are $K_{0c}=1/\sqrt{1+U_0/\pi v^0_F}$ and $K_{c}=1/\sqrt{1+2U_1/\pi v_F}$,
and the renormalized velocities are $\tilde{v}_{Fc}^0=v_F^0/K_{c0}$ and $\tilde{v}_{Fc}=v_F/K_c$. 
The coupling between $\tilde{\phi}_{0c}$ and $\tilde{\phi}_{c}$ modes is given by 
$\lambda=U'\sqrt{2K_{0c}K_c}/\pi$.
The particular  band structure of the thinest SWNTs leads to the existence of the two charge modes which are coupled
by the Coulomb interaction.

The Hamiltonian  (\ref{5}) is bilinear and can be diagonalized  
taking the free-boson form with 
two characteristic  velocities 
$
v_{\pm c}^{\; 2}=[ \tilde{v}_F^{0\;2} +\tilde{v}_F^{\;2}\pm
\sqrt{
(\tilde{v}_F^{0\;2} -\tilde{v}_F^{\;2})^2 
+4\lambda^2\tilde{v}_F^{0}\tilde{v}_F}
 ]\big/2
$ \cite{bogoliubov}.
The solution is real only if $v_{\pm c}^{\;2}\geq 0$, i.e.
 $\tilde{v}_F^{0}\tilde{v}_F\geq \lambda ^2$.
For large   $\lambda$  
the system is unstable toward the long-wave length density fluctuations.
This kind of instability is known as the Bardeen-Wentzel instability (BWI), 
originally discussed in the case of  the 1-d electrons 
coupled to phonons \cite{Wentzel51,Loss94}. 
The BWI for two electrostatically coupled Luttinger liquids was also discussed in Ref. \cite{Loss94}. 
In a proximity  to the BWI
but still on the metallic side, the repulsive interband interaction 
can lead to a  retarded intraband attraction and the electrons 
 may  form Cooper pairs. 
This is a possible microscopic mechanism responsible for 
developing the SCFs in the $4\AA$-SWNTs.

In the neutral sector, the Hamiltonian takes the simple free-boson form
with four characteristic velocities 
$\tilde{v}_{Fs}^0=v_F^0/K_{0s}$ and $\tilde{v}_{F\nu}=v_F/K_{\nu}$, where $K_{0s}=1$ and $K_{\nu}=1$,
and $\nu=s,f$ or $x$.
We have left the Luttinger liquid parameters ($K_{0s}$ and $K_{\nu}$) explicitly
because they are expected to flow due to a renormalization   
caused by backscattering processes.

The one-particle correlation functions at zero temperature decay with a distance  
$\langle \psi_{rj\sigma}(x) \psi_{rj\sigma}^{\dagger}(0) 
\rangle \sim x^{-\alpha_j} $, where the interaction dependent exponents 
$
\alpha_0=K_{0s}/4+1/4K_{0s} + K_{0c}A_0/4 + B_{0}/4K_{0c}$, and
$
\alpha_1=\alpha_2=
K_s/8+1/8K_s+K_f/8+1/8K_f+K_x/8+1/8K_x+K_cA/8+B/8K_c$.
The coefficients $A$s and $B$s are given by certain combination of the effective
 velocities and Bogoliubov parameters \cite{coefficients}.

Deviations form $\alpha_j=1$ indicate  the Luttinger liquid behavior and can be experimentally detected
by measuring the tunneling conductance through the SWNT \cite{LL}. 
Each transport  channel ($j$) contributes in parallel to the total conductance $G$ by
its partial conductance $G_j(T,V_{sd})=T^{\alpha_j-1}F(eV_{sd}/kT)$, where $T$ is the temperature,
$V_{sd}$ is an applied bias voltage, and $F(z)$ is a known universal function \cite{LL}. 
We predict the 
following form of the conductance
$ G(T,V_{sd})=(T^{\alpha_0-1}+2T^{\alpha_1-1})F(eV_{sd}/kT)$ in the thinest zigzag SWNTs.

Two-particle correlation functions at zero temperature decay according to 
 the power law $ x^{-\beta_j}$ (with $\beta_j=2$ in the noninteracting case).
The slowest decaying correlation function indicates which kind of fluctuations dominates in the system.
We calculate  the exponents $\beta_j$ corresponding to: a singlet superconducting order parameter (SSC),
a triplet superconducting order parameter (TSC), a charge/spin density wave order parameter 
(CDW/SDWz), and 
a spiral density wave order parameter (SDWs), 
with the nesting vectors $Q_j=2k_F^j$ for the CDW/SDWz and SDWs cases ($k_F^j$ is the Fermi vector 
in the corresponding band).
The results for the exponents of the intraband correlation functions are summarized in the 
Table I.
The interband two-particle correlation functions decay much faster with $x$ and are not relevant here.

\begin{table}
\caption{\label{table} Exponents for the two-particle correlation functions.}
\begin{ruledtabular}
\begin{tabular}{ccc}
 & bands-$00$ & bands-$11$($22$)\\
\hline
SSC & $K_{0s}+\frac{B_0}{K_{0c}}$ & $\frac{K_s}{2}+\frac{K_x}{2}+\frac{1}{2K_f}+\frac{B}{2K_c}$\\
TSC & $\frac{1}{K_{0s}}+\frac{B_0}{K_{0c}}$ & $\frac{1}{2K_s}+\frac{1}{2K_x}+\frac{1}{2K_f}+\frac{B}{2K_c}$\\
CDW/SDWz & $K_{0s}+K_{0c}A_0$ & $ \frac{K_s}{2}+\frac{K_x}{2}+\frac{K_f}{2}+\frac{K_c}{2}A $\\
SDWs & $\frac{1}{K_{0s}}+K_{0c}A_0$ & $ \frac{1}{2K_s}+\frac{1}{2K_x}+\frac{K_f}{2}+\frac{K_c}{2}A $
\end{tabular}
\end{ruledtabular}
\end{table}

\begin{figure}
\includegraphics [clip,width=7.cm,angle=-00]{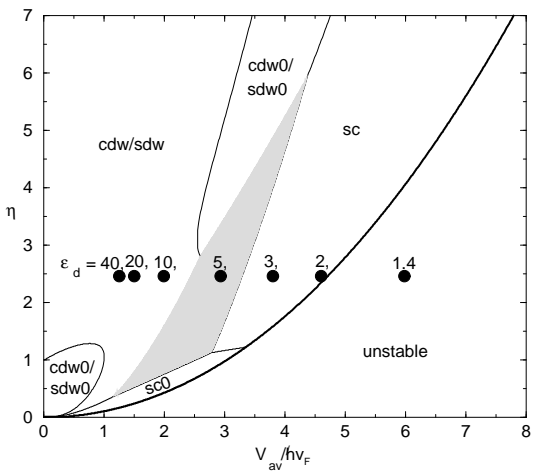}
\caption{ 
The phase diagram for the thinest zigzag  carbon nanotubes in a parameter space given by 
 the average  bare interaction 
$V_{\rm av}/\hbar v_F$ and the  ratio $\eta=v^0_F/v_F$ of the Fermi velocities. 
The sector of the diagram with the dominant superconducting fluctuations 
in the $j=1$ and $2$ ($j=0$) bands is 
marked by SC (SC0) and the portion with the dominant charge/spin density wave fluctuations  
in the $j=1$ and $2$ ($j=0$) bands is indicated by CDW/SDW (CDW0/SDW0).
The gray area shows  the metallic phase where all of the two-particle fluctuations are suppressed. 
The thick solid line points out the Bardeen-Wentzel instability.
Large dots represent  the averages for the bare interactions (\ref{potential}) with different 
 $\epsilon_d$.  } 
\label{fig2}
\end{figure}

In Fig. \ref{fig2} we present a phase diagram at zero temperature in the parameter space defined by 
the average bare interaction $V_{\rm av}$ (in units of $\hbar v_F$) and the  ratio $\eta=v_F^0/v_F$.
For large $\eta$ ($\gtrsim 6$) and small $V_{\rm av}/v_F$  the density wave fluctuations (DWFs) in the 
$j=1$ and $2$ bands
appear to have the longest decaying length.
For moderate $V_{\rm av}/\hbar v_F$,  the DWFs dominate in the $j=0$ band
 and finally there is a wide part 
of the diagram where the SCFs develop  in the $j=1$ and $2$ bands.
This phase is terminated by the line where the BWI occurs.
For moderate $\eta$, an additional  phase appears  where all kind of the two-particle 
fluctuations are suppressed
 (the gray area in Fig. \ref{fig2}).
For small $\eta$ and $V_{\rm av}/\hbar v_F$,  the DWFs evolve in the $j=0$ band 
 and  for larger $V_{\rm av}/\hbar v_F$, close to the BWI line, 
the SCFs are amplified in the $j=0$ band.

The SCFs with the longest  decaying length develop in a band with a smaller
renormalized velocity.
In other bands, with a larger renormalized velocity,  the SCFs are also amplified  (with $\beta_j<2$). 
However, when the SCFs are enhanced,  the DWFs are  suppressed (with $\beta_j>2$) in all of the bands.
Since $K_{0s}=K_{\nu}=1$, the exponents in the SSC and the TSC correlation functions are the same.
This is also the case for the CDW/SDWz and the SDWs correlation functions.
Further analysis, including the renormalization of the Luttinger liquid parameters
due to backscattering processes, would resolve which 
particular correlation function has the longest  decaying length.

We estimate the average value $V_{\rm av}$ of the bare interaction  for the different dielectric constants 
$\epsilon_d$ of the zeolite host. 
We take the radius of the SWNT $a=2.1\AA$ and the radius of the 1-d cavity in the host
$R=3.7\AA$ \cite{Qin00,Tang01}.
The Fermi velocities are taken $v^0_F=6.9\cdot 10^5\;m/s$ and $v_F=2.8\cdot 10^5\;m/s$ \cite{Blase94}, which 
give the ration $\eta=2.46$. 
The dielectric constant for an isolated SWNT is $\epsilon_1=1.4$ \cite{Egger97}.
The average potential decreases with increasing $\epsilon_d$ as is shown by dots going from the right to the left 
 in Fig.\ref{fig2}.
For $\epsilon_d=\epsilon_1=1.4$, our estimation predicts that the metallic phase is unstable.
For $\epsilon_d=2$ and $3$, the system is metallic and dominated by the SCFs.
For $\epsilon_d=5$, we find that all two-particle fluctuations are suppressed.
For $\epsilon_d=10,20$ and $40$, the DWFs appear to be  dominant.

Dielectric properties of the host can change the physical properties
of the SWNT. 
Creating and investigating  electrical properties of the SWNTs in zeolites
with different dielectric constants would be a direct verification of 
our theory.
For the zeolite AlPO$_4$-5 with the dielectric constant $\epsilon_d\approx 2-4$, 
our theory predicts that  the SCFs are amplified according to the experiment \cite{Tang01}.
The dielectric constant of this zeolite after filling it in with the 
water is $\epsilon_d\approx 40$ \cite{zeolite}.
If one can still synthesis the SWNTs there, they should develop the DWFs. 

In conclusion, fabricating the SWNTs inside the zeolite hosts with different dielectric constants
 would provide a rare opportunity to control the electron-electron interaction 
and to examine the  different phases in such  1-d systems.

It is a pleasure to  thank Prof. D. Vollhardt for discussion. 
Also information from Prof. P. Sheng on the experiment 
\cite{Tang01} are kindly acknowledged.
This work was supported by the Alexander von
 Humboldt-Foundation.




\begin{references}
\bibitem{comment0}
Due to the time reversal symmetry the matrix elements $V_{1111}=V_{2222}$.
Additionally, in the $q\rightarrow 0$ limit the fast-oscillating terms  $V_{1122}$ can be neglected and
the polarization functions $\Pi_{01}\approx \Pi_{02}\approx \Pi_{12}\approx 0$. 
Also, because of the degeneracy between the $j\neq 0$ bands, $V_{1212}=V_{1111}$.
\bibitem{comment1}
Formally, the potential $U_{ij}(x)$  should be slightly different from that given in the text
because in deriving the effective Hamiltonian (\ref{2}) one should terminate the renormalization 
of the interaction at some finite 
energy  $E_c >0$. 
However, the difference should be small and can be disregarded in the following.
\bibitem{bogoliubov}
The coefficients in the Bogoliubov transformation diagonalizing (\ref{5}) are given by
$
\alpha^2 (\beta^2)=[ 1\pm|\tilde{v}_F^{0\;2} -\tilde{v}_F^{\;2}|/\sqrt{
 (\tilde{v}_F^{0\;2} -\tilde{v}_F^{\;2})^2 
+4\lambda^2\tilde{v}_F^{0}\tilde{v}_F}]\big/ 2
$.
\end{references}
\end{document}